\documentclass{article}
\usepackage{amsmath}
\usepackage{amssymb}
\usepackage{graphicx}
\usepackage{epstopdf}
\begin{document}

\title{Quantum cosmological intertwining: Factor ordering and boundary conditions from hidden symmetries}
\author{T. Rostami\thanks{email:t\_rostami@sbu.ac.ir}\\{\small Department of Physics, Shahid Beheshti University G. C., Evin, Tehran 19839, Iran}\,\,\and\,\, \\ S. Jalalzadeh\thanks{email: s-jalalzadeh@sbu.ac.ir;
shahram.jalalzadeh@unila.edu.br} \\{\small Department of Physics, Shahid Beheshti University G. C., Evin, Tehran 19839, Iran}\\{\small Federal University of Latin-America Integration, Technological Park of Itaipu,}\\ {\small PO box 2123, Foz do Igua\c{c}u-PR, 85867-670, Brazil}\,\, \and\,\, \\
P. V. Moniz\thanks{e-mail: pmoniz@ubi.pt}\\{\small Centro de Matem\'{a}tica e Aplica\c{c}\~{o}es- UBI, Covilh\~{a}, Portugal},\\
{\small Departmento de F\'{\i}sica, Universidade da Beira Interior, 6200 Covilh\~{a}, Portugal}}
\date{\today}
\date{\today}
\maketitle

\begin{abstract}
We explore the implications of hidden symmetries present in a particular quantum cosmological setting, extending the results reported in
\cite{10,11}. In more detail, our case study is constituted by a spatially closed Friedmann-Lema\^{\i}tre-Robertson-Walker
universe, in the presence of a conformally coupled scalar field. The $su(1,1)$ hidden symmetries of this model, together with the
Hamiltonian constraint, lead to the gauge invariance of its corresponding Bargmann indices. We subsequently show that some factor-ordering
choices can be related to the allowed spectrum of Bargmann indices and hence, to the hidden symmetries. Moreover, the presence of those
hidden symmetries also implies a set of appropriate boundary conditions to choose from. In summary, our results suggest that factor ordering and boundary conditions can be intertwined when a quantum cosmological model has hidden symmetries.\\

\end{abstract}

\section{INTRODUCTION}
An important issue in theoretical cosmology is the choice of an initial condition for our Universe to evolve from. Although some features of the observed
Universe are explained by the hot big bang model, it faces the problem of describing properties such as its spatial flatness and isotropy, plus
the
origin of density fluctuations. The inflationary scenario \cite{1} proposes to solve these problems
with adequate density fluctuations being obtained if the matter fields start in a particular quantum state \cite{Boubekeur:2015xza}. Notwithstanding the success so far achieved by the inflationary paradigm \cite{A1}, in order to have a complete understanding of the
present observable state of the Universe, the initial condition problem must be addressed.

The Wheeler-De Witt (WDW) equation plays an important role in quantum cosmology \cite{2}, determining the wave function of
the Universe. A framework that has been usually employed is the Hamiltonian formulation of general relativity introduced by Arnowitt,
Deser and Misner (ADM), by means of a decomposition of the spacetime manifold \cite{3}. Nevertheless, there are pertinent
technical challenges \cite{4}.

The WDW equation has many solutions: to single out an appropriate quantum state, a sensible
procedure assisting in the selection of  boundary or initial conditions is needed \cite{5}. Different approaches
have been suggested to address the problem of boundary conditions, but these were not conveyed as part of a dynamical law \cite{Hartle:1996xa}. Let us elaborate more on this. The motivation
behind the no-boundary approach \cite{Hawking:1983hj} was to surmount the initial singularity by determining the wave function of the Universe
through a path integral over compact Euclidean geometries. The tunneling proposal \cite{Vilenkin:1982de} determines the wave function to be
bounded everywhere and includes only outgoing modes at the singularity. Two other proposals, the vanishing of the wave
function or its derivative with respect to the scale factor at the classical singularity \cite{dewitt, 6scalar}, have also
been used to specify the wave function of the Universe.

In addition, specific settings in quantum cosmology often require factor-ordering choices within the WDW equation to be investigated.
It has also been stated, however, that the factor-ordering question is not very important to the theory as
a
whole \cite{dewitt,Vil}, namely, from a semiclassical perspective \cite{Bojowald:2014ija, 6a}. Nevertheless, it has been claimed that different operator orderings
\cite{6a,6,7}
can be related to a chosen boundary condition \cite{7}. Unfortunately, there is no general argument on how to resolve this issue and therefore, proposals given in this regard
are
somewhat phenomenological \cite{dewitt}. In this context, it may be relevant to point to the following. The
classical theory of general relativity is invariant under the group Diff(${\mathcal M}$) of diffeomorphism of the
spacetime manifold $\cal M$,  which leads to the problem of observables \cite{Bojowald:2014ija}. On the other hand, the identification of a
dynamical
observable is related to the issue of time \cite{Isham:1992ms,8}.
Let us be more concrete. The observables of a theory, according to Dirac \cite{9}, are those
quantities which have vanishing Poisson brackets at the classical level. Hence, at the quantum level, they satisfy the
appropriate quantum commutators in the presence of constraints. In the ADM formalism, several constraints emerge: in particular, the primary constraints associated to the
canonical conjugate momentum of the lapse function and shift vector. Subsequently, secondary constraints can be retrieved, namely the Hamiltonian or momentum constraints. Finally,
all
Dirac observables are time independent.

Because the algebra of the theory is specified by the constraints, it may be thought that there is a relation
between the choice of the boundary condition (and then of the factor ordering \cite{6a,6,7}) and thus, the Dirac observables of the theory.
More
precisely, in Ref. \cite{10} the Hamiltonian of a closed Friedmann-Lama\^{\i}tre-Robertson-Walker (FLRW) universe filled with either dust or radiation was found to be equivalent to the one-dimensional harmonic oscillator. The hidden symmetry of that model $su(1,1)$,
together with the gauge invariance of the Bargmann index with values $\{\frac{1}{4},\frac{3}{4}\}$, split the underlying
Hilbert space into two disjoint invariant subspaces; each corresponding to a different choice of boundary condition. In a
similar procedure, in Ref. \cite{11} the $u(1,1)$ hidden symmetries of the nonminimally coupled scalar field in a
spatially flat universe, led to the Hamiltonian of that model being described by a two-mode realization of the $su(1,1)$
algebra, which induces a degenerate Bargmann index. Therein, the scale factor duality of that model together with time
reversal, allows us to specify the appropriate boundary conditions.
Therefore, from the results conveyed in \cite{10,11}, hidden symmetries present in the Hamiltonian, in addition to the
minisuperspace
symmetries of the model under investigation, suggest a process from which to select boundary
conditions.

Herein, we extend the scope of \cite{10,11} towards a closed FLRW model nonminimally coupled to a scalar field.
The novel contribution brought in this paper is to provide a concrete procedure to relate, in an intertwined manner, factor ordering and boundary conditions making use of the presence of a specific hidden symmetry and by means of a Dirac observable algebra. In Sec II, we present the model that assists in our investigation. The
quantization of the model and a boundary condition discussion (in view of the retrieved Dirac observables as well as concrete factor-ordering choices) are analyzed in detail in Sec
III. A summary and discussion of our results is presented in Sec IV.

\smallskip
\section{CLASSICAL SETTING}
The action of general relativity nonminimally coupled to a scalar field is
\begin{eqnarray}\label{0,1}
{\cal{S}}=\int d^{4}x\sqrt{-g}\left[\left(\frac{1}{2\kappa}-\frac{\zeta}{2}\phi^{2}\right)R-\frac{1}{2}\partial_{\mu}\phi
\partial^{\mu}\phi-V(\phi)\right],
\end{eqnarray}
where $\kappa=8\pi G$ with $G$ being Newton's gravitational constant and $\zeta$ is a dimensionless coupling constant. One of the main reasons to include the nonminimal coupling in the action is that at the quantum level, quantum corrections to the scalar field theory lead to the nonminimal
coupling: the scalar field theory in curved spacetime becomes renormalizable in the case of a nonminimal
coupling \cite{12}. Furthermore, the recent Planck data \cite{Boubekeur:2015xza} suggests for the early Universe a stage where a nonminimal coupling may have had a suitable contribution. Different values of the nonminimal coupling have been adopted \cite{12,15}. In metric theories of gravity, $\zeta=0$ (minimal coupling) or
$\zeta=\frac{1}{6}$ (conformal coupling) have been frequently employed
\cite{019}. In grand unified theories (GUTs), $\zeta$ depends on a renormalization group
parameter $\tau$, and $\zeta(\tau)$ converges to $1/6$, $\infty$, or $\zeta_{0}$ as $\tau\rightarrow \infty$
\cite{020}. In the standard model, the Higgs fields have been associated to either $\zeta\geq 1/6$ or $\zeta\leq 0$ \cite{021}.
Herein, we adopt the special value $\zeta=\frac{1}{6}$ with $V=0$ which  makes the physics of $\phi$
conformally invariant \cite{19,20}.

In addition, we use a closed FLRW geometry, with the line element
\begin{eqnarray}\label{0,2}
ds^{2}=-N(t)^{2}dt^{2}+a(t)^2\left(\frac{dr^{2}}{1-r^2}+d\Omega_{3}^{2}\right).
\end{eqnarray}
Introducing the new variable $\widetilde{\phi}=\frac{al_{p}\phi}{\sqrt{2}}$, where $l_{p}$ is the Planck length $(8\pi
G=3{l_p}^{2})$, the Hamiltonian for our model, corresponding to a two-dimensional minisuperspace, $\{a,\widetilde{\phi} \}$, is
\begin{eqnarray}\label{0,3}
H=N\left[-\frac{{\Pi_{a}}^{2}}{4a}+\frac{{\Pi_{\widetilde{\phi}}}^{2}}{4a}-a+\frac{{\widetilde{\phi}}^{2}}{a}\right],
\end{eqnarray}
where $\Pi_{a}=\frac{-2\dot{a}a}{N}$, $\Pi_{\widetilde{\phi}}=\frac{-2\dot{\widetilde{\phi}}a}{N}$ and $\Pi_{N}=0$ are
the
canonical momenta conjugate to $a$, $\widetilde{\phi}$, and $N$, respectively. In the presence of the primary constraint $\Pi_{N}=0$,
the Hamiltonian can be generalized by adding to it this primary constraint multiplied by arbitrary functions of time
$\xi$. Then, the total Hamiltonian will be
\begin{eqnarray}\label{0,4}
{\cal{H}}_{T}=N\left[-\frac{{\Pi_{a}}^{2}}{4a}+\frac{{\Pi_{\widetilde{\phi}}}^{2}}{4a}-a+\frac{{\widetilde{\phi}}^{2}}{a}\right]+\xi
\Pi_{N}.
\end{eqnarray}
The primary constraint must be satisfied at all times and therefore,
\begin{equation}\label{0,13}
{\dot{\Pi}_{N}}=\{\Pi_{N},{\cal{H}}_{T}\}\approx0,
\end{equation}
which leads to the secondary (Hamiltonian) constraint,
\begin{equation}\label{0,14}
H=N\left[-\frac{{\Pi_{a}}^{2}}{4a}+\frac{{\Pi_{\widetilde{\phi}}}^{2}}{4a}-a+\frac{{\widetilde{\phi}}^{2}}{a}\right]\approx0.
\end{equation}
The existence of constraint (\ref{0,14}) means that there are some degrees of freedom which are not physically
relevant.
Hence, we can fix the gauge as $N=a$. Then, the Hamiltonian can be readily written as
\begin{equation}\label{0,15}
H=\left[-\frac{{\Pi_{a}}^{2}}{4}+\frac{{\Pi_{\widetilde{\phi}}}^{2}}{4}-a^2+{\widetilde{\phi}}^{2}\right]\approx0.
\end{equation}
\subsection{Reduced phase space and observables}
According to Dirac, an observable is a function on the phase space which has weakly vanishing Poisson brackets with
the
first-class constraints. A phase space function is a first-class constraint if its Poisson bracket with all constraints
weakly vanishes.
In particular, general relativity is invariant under the group of diffeomorphisms, the Hamiltonian can be expressed as
a sum of constraints, and any observable must commute with these constraints.

Therefore, in order to find the corresponding gauge invariant observables of our model in (\ref{0,1}) -(\ref{0,2}), we
define
the complex valued functions $S=\{K_{0},K_{\pm},J_{0},J{\pm} \}$ on the unconstrained phase space $\Gamma$ in
${\mathbb{R}}^{6}$. More concretely, $\{K_{0},K_{\pm}\}$ are the complex valued functions for the gravitational sector of the Hamiltonian, and are defined as
\begin{eqnarray}\label{4,4}
\begin{cases}
K_{0}=\frac{1}{4}\left(a^2+\Pi_{a}^{2}\right),\\
K_{\pm}=\frac{1}{4}\left(a^2-\Pi_{a}^{2}\mp i\left(a\Pi_{a}+\Pi_{a}a\right)\right),
\end{cases}
\end{eqnarray}
with the following closed Poisson algebra
\begin{eqnarray}\label{4,6}
\begin{array}{lll}
\left\{K_{0},K_{\pm}\right\}=\mp iK_{\pm},\hspace{.3cm}
\left\{K_{+},K_{-}\right\}=2iK_{0}.
\end{array}
\end{eqnarray}
Moreover, we have the complex valued functions $\{J_{0}, J_{\pm}\}$ for the scalar field part as
\begin{eqnarray}\label{4,5}
\begin{cases}
J_{0}=\frac{1}{4}\left(\Pi_{\widetilde{\phi}}^{2}+{\widetilde{\phi}}^{2}\right),\\
J_{\pm}=\frac{1}{4}\left(\widetilde{\phi}^2-\Pi_{\widetilde{\phi}}^{2}\mp
i\left(\widetilde{\phi}\Pi_{\widetilde{\phi}}+\Pi_{\widetilde{\phi}}\widetilde{\phi}\right)\right).
\end{cases}
\end{eqnarray}
They satisfy the following closed algebra
\begin{eqnarray}\label{4,7}
\begin{array}{lll}
\left\{J_{0},J_{\pm}\right\}=\mp iJ_{\pm},\hspace{.3cm}
\left\{J_{+},J_{-}\right\}=2iJ_{0}.
\end{array}
\end{eqnarray}
Using (\ref{4,4}) and (\ref{4,5}) the Hamiltonian constraint (\ref{0,15}) becomes
\begin{eqnarray}\label{4,2}
{\ H}=4\left(K_{0}-J_{0}\right)\approx 0,
\end{eqnarray}
which shows that $J_{0}$ and $K_{0}$ are not independent. Since the Poisson brackets of all above observables of the phase space and the Hamiltonian
vanish, i. e., $\{H,K_{0}\}=\{H,K_{\pm}\}=\{H,J_{0}\}=\{H,J_{\pm}\}=0$, their values on the constraint surface are constants of motion. Furthermore, for the gravitational part, if we define the central element of the algebra as
\begin{eqnarray}\label{4,3}
K^{2}:=\frac{1}{2}K_{0}^{2}- \frac{1}{4}\left(K_{+}K_{-}+K_{-}K_{+}\right),
\end{eqnarray}
then, by inserting the definitions displayed in expressions (\ref{4,4}) into Eq. (\ref{4,3}), we can easily show that on the constraint surface
$H\approx 0$, the $K$'s are not algebraically independent but satisfy the
identity
\begin{eqnarray}\label{4,13}
K^{2}=-\frac{3}{16}.
\end{eqnarray}
Similarly, for the scalar field part, we introduce
\begin{eqnarray}\label{4,11}
\begin{array}{lll}
J^{2}:=\frac{1}{2}J_{0}^{2}- \frac{1}{4}\left(J_{+}J_{-}+J_{-}J_{+}\right).
\end{array}
\end{eqnarray}
Using definitions (\ref{4,5}) in Eq.(\ref{4,18}), the $J$'s satisfy the identity
\begin{eqnarray}\label{4,18}
J^{2}=-\frac{3}{16}.
\end{eqnarray}
Obviously, $K^{2}$ and $J^{2}$ have strongly vanishing Poisson brackets with the Hamiltonian and their values are constant
of motion. Note that by means of the three constraints (\ref{4,2}), (\ref{4,18}), and (\ref{4,13}), three of the $S=\{K_{0},K_{\pm},J_{0},J{\pm} \}$ are independent on the phase space.

In this paper, we follow the argument that the (Dirac) observables are characterized by having weakly vanishing Poisson brackets with first-class constraints \cite{9}. Another approach that we have not considered in our paper has been promoted in \cite{pitts}, by means of which observables, in general, should have vanishing Poisson brackets with the gauge generators. However, by assuming Dirac's suggestion that gauge generators are the first-class constraints, in our case study the appropriate gauge generator to be contemplated for the action (\ref{0,1}) reduces to (\ref{0,4}), and the observables determined here in our paper are consistent within the approach we took.

\smallskip

\section{QUANTUM COSMOLOGY}
A quantum state\footnote{In our paper, we take $\hbar=1$.} for the FLRW universe (\ref{0,1}) -(\ref{0,2}) can be
obtained from the WDW equation. With $H\Psi(a,\widetilde{\phi})=0$ and
$\Pi_{a}^{2}\equiv-a^{-q}\frac{\partial}{\partial_{a}}\left(a^{q}\frac{\partial}{\partial_{a}}\right)$, $\Pi_{\widetilde{\phi}}=-i\frac{\partial}{\partial_{\widetilde{\phi}}}$, we write the WDW equation as
\begin{eqnarray}\label{1,0}
\left[a^{-q}\frac{\partial}{\partial_{a}}\left(a^{q}\frac{\partial}{\partial_{a}}\right)-a^{2}-\left
\{\frac{\partial^{2}}{\partial \widetilde{\phi}^{2}}-{\widetilde{\phi}^{2}}\right \}\right]\Psi(a,\widetilde{\phi})=0.
\end{eqnarray}
An important point to note is a factor-ordering ambiguity, by means of the  power $q$ in the first term,
which can arise through the canonical quantization procedure \cite{Bojowald:2014ija,6a}. One factor-ordering choice (among several possibilities) is the natural ordering, which induces a Laplace-Beltrami operator in the minisuperspace \cite{dewitt}. The existence of arbitrary possible choices for the factor ordering is a relevant issue in quantum cosmology. In what follows, we will adopt the general factor-ordering introduced above in (\ref{1,0}), and we will show that within this specific factor-ordering structure, some choices can be selected for the algebra of observables of the model. Then they can be subsequently associated with a boundary condition selection, extracted from an analysis of the hidden symmetries present in the model.
Rewriting the wave function as $a^{\frac{-q}{2}}\Psi(a,\widetilde{\phi})$, the WDW equation (\ref{1,0}) simplifies to
\begin{eqnarray}\label{1,1}
\left[\frac{\partial^{2}}{\partial a^{2}}-\left\{\frac{\partial^{2}}{\partial
\widetilde{\phi}^{2}}-\widetilde{\phi}^{2}\right\}-a^{2}-\frac{\beta}{a^2}\right]\Psi(a,\widetilde{\phi})=0,
\end{eqnarray}
where $\beta=\frac{q(q-2)}{4}$ is a parameter representing the operator ordering ambiguity in the first term of Eq. (\ref{1,0}) \cite{21}.
The WDW equation for a conformally coupled FLRW cosmology is a very special case for which we can separate completely the scalar
field part from the gravitational sector, i. e., $H=H_{a}\oplus\ H_{\widetilde{\phi}}$.
For the conformal scalar field part, with a separation constant $E_{n}$, we have
\begin{eqnarray}\label{1,2}
\left[-\frac{\partial^{2}}{\partial
\widetilde{\phi}^{2}}+\widetilde{\phi}^{2}\right]\Phi_n(\widetilde{\phi})=E_{n}\Phi_n(\widetilde{\phi}).
\end{eqnarray}
The solution to the above equation is
\begin{eqnarray}\label{1,3}
\begin{cases}
\Phi_{n}(\widetilde{\phi})={\mathcal N}_{n}H_{n}(\sqrt{2}\widetilde{\phi})e^{-\frac{\widetilde{\phi}^{2}}{2}},\\
E_{n}=2n+1,~~~~ n=0,1,2,...,
\end{cases}
\end{eqnarray}
where $H_{n}$ is an Hermite polynomial of order $n$, and $\mathcal N_{n}$ is a integration constant. For the gravitational sector
we
have
\begin{eqnarray}\label{1,4}
\left[-\frac{d^{2}}{da^{2}}+a^{2}+\frac{q(q-2)}{4a^{2}}\right]\psi_{n^{\prime}}(a)=\bar{E}_{n^{\prime}}\psi_{n^{\prime}}(a),
\end{eqnarray}
The solution to the above equation is \cite{22, Ha}
\begin{eqnarray}\label{1,5}
\begin{cases}
\psi_{n^{\prime}}^\gamma(a) = {\cal N}_{n^{\prime}}^\gamma\ a^{\gamma+1} e^{-\frac{1}{2}a^2} L_{n^{\prime}}^{\gamma+\frac{1}{2}}(a^2), \\
\bar{E}_{n^{\prime}}^\gamma=4(n'+\frac{1}{4}-\frac{\gamma}{2}),\qquad n^{\prime} = 0, 1, 2, \ldots,
\end{cases}
\end{eqnarray}
where $L_{n'}^{\gamma+\frac{1}{2}}$ as generalized Laguerre polynomials
of degree $n'$ \cite{abramowitz},
\begin{equation}
  {\cal N}_{n^{\prime}}^\gamma = (-1)^{n^{\prime}} \left(
  \frac{n^{\prime}!}{\Gamma(n^{\prime}+\gamma+\frac{3}{2})}\right)^{\frac{1}{2}},  \label{eq:HO-norm}
\end{equation}
is a normalization coefficient and $\gamma=\frac{1}{2}(-1\pm|1-q|)$. It is obvious that for the specific choice of $\gamma=0$ (equivalent to $q=0$ or $q=2$) the WDW equation consists of two harmonic
oscillators
with opposite signs \cite {22} which are regular everywhere.
The Hamiltonian constraint for the conformally coupled scalar field (\ref{0,14}) leads to
\begin{eqnarray}\label{1,8}
2n-4n^{\prime}=q.
\end{eqnarray}
\subsection{Hidden symmetries and boundary conditions}
In order to determine the wave function of the Universe, given the mathematical nature of the WDW equation, boundary conditions must be imposed. A singularity is present at $t=0$ and physically relevant configurations require that the scale factor, $a$, is positive. Hence,
the configuration space for the gravitational sector is the half-line $(0,\infty)$. Thus, the corresponding Hilbert space related to $H_a$ is $L^2(0,\infty)$, with the following inner product
\begin{eqnarray}\label{inner}
\langle\psi_{1},\psi_2\rangle=\int_{0}^{\infty}\psi^{\dagger}_{1}(a)\psi_{2}(a)da.
\end{eqnarray}
However, the operator $H_{a}=-\frac{d^{2}}{d a^{2}}+a^{2}+\frac{q(q-2)}{4a^2}$ defined on $C_{0}^{\infty}(0,\infty)$ is not necessarily self-adjoint. To have a self-adjoint Hamiltonian, it is necessary to have simultaneously $\psi(0^+)=0$ and $\frac{d\psi}{da}(0^+)=0$, or the domain of $H_a$ should be restricted to the domains \cite{Lemos},
\begin{eqnarray}\label{l}
D_{\alpha}=\{\psi\in H_{a}^2(0,\infty)|\frac{d\psi}{da}(0^+)=\alpha\psi(0^+)\},
\end{eqnarray}
where $\alpha\in\mathbb{R}$ and $H_a^2(0,\infty)$ denote the Sobolev space with the wave functions $\psi\in L^2(0,\infty)$ with $\psi\in C^1(0,\infty)$, continuous $\frac{d\psi}{da}$,  $\frac{d^2\psi}{da^2}\in L^2(0,\infty)$, and $H_a[\psi]\in L^2(0,\infty)$.
Note that for simultaneously vanishing $\psi(0^+)$ and $\frac{d\psi}{da}(0^+)$, the equation $\frac{d\psi}{da}(0^+)=\alpha\psi(0^+)$ will be trivial. In \cite{6scalar}, the simple cases of $\alpha=0$ and $\alpha=\infty$ have been used. Moreover, it is argued in \cite{24} that this arbitrary constant would be a new fundamental physical constant and in order to avoid such a constant, $\alpha$ is required to be zero. However, $\alpha$ can be determined in the context of a hidden dynamical symmetry being present.
The lowering and raising operators for the WDW equation (\ref{1,4}) can be
built using a factorization method \cite{l'Yi:1995bg}. Let us start with the WDW equation (\ref{1,4}) and
rewrite it as a one-dimensional Schr\"odinger equation
\begin{eqnarray}\label{r1}
\begin{cases}
H^\gamma\psi^\gamma_{n'}=\bar E_{n'}^\gamma\psi_{n'}^\gamma,\\
H^\gamma:=-\frac{d^2}{da^2}+a^2+\frac{\gamma(\gamma+1)}{a^2},
\end{cases}
\end{eqnarray}
where $\gamma=\frac{1}{2}(-1\pm |1-q|)$. Introducing the first-order differential
operators
\begin{eqnarray}\label{r2}
\begin{cases}
C_\gamma:=\frac{d}{da}+a+\frac{\gamma}{a},\\
C^\dagger_\gamma:=-\frac{d}{da}+a+\frac{\gamma}{a},
\end{cases}
\end{eqnarray}
we obtain the following supersymmetric partner Hamiltonians \cite{A1,Cooper:1994eh}:
\begin{eqnarray}\label{r3}
\begin{cases}
h_+:=C_\gamma C^\dagger_\gamma=H^\gamma+2\gamma-1,\\
h_-:=C_\gamma^\dagger C_\gamma=H^{\gamma-1}+2\gamma+1.
\end{cases}
\end{eqnarray}
Then, the Hamiltonians $h_+$ and $h_-$ have the same energy spectrum except
the ground state of $h_+$
\begin{eqnarray}\label{r4}
\begin{cases}
h_+\psi_{n'}^\gamma=(\bar E_{n'}^\gamma+2\gamma-1)\psi_{n'}^\gamma,\\
h_-\psi_{n'-1}^{\gamma-1}=(\bar E^{\gamma-1}_{n'-1}+2\gamma+1)\psi_{n'-1}^{\gamma-1}=(\bar
E_{n'}^\gamma+2\gamma-1)\psi_{n'-1}^{\gamma-1}.
\end{cases}
\end{eqnarray}
This symmetry is called shape-invariant symmetry \cite{Cooper:1994eh}. The shape-invariant
condition (\ref{r4}) is equivalent to
\begin{eqnarray}
C_\gamma C_\gamma^\dagger-C_{\gamma-1}^\dagger C_{\gamma-1}=4.
\end{eqnarray}
We see that changing the order of operators $C_\gamma^\dagger$ and $C_\gamma$
leads to a shift in the value of $\gamma$. The above discussion shows that
the different factor-orderings of the WDW equation are related through shape-invariance features.
It is well known that the shape-invariant potentials are easy to deal with if
lowering and raising operators, just as for the harmonic oscillator, are employed.
However, as the commutator of $C_\gamma$ and $C_\gamma^\dagger$ does not
yield a constant value, namely,
\begin{eqnarray}\label{r5}
[C_\gamma,C^\dagger_\gamma]=2\left(1-\frac{\gamma}{a^2}\right),
\end{eqnarray}
these operators are not suitable. To establish a suitable algebraic structure,
according to \cite{l'Yi:1995bg}, we assume that replacing $\gamma$ with $\gamma-1$
in a given operator can be achieved with a similarity transformation,
\begin{eqnarray}\label{r6}
T\mathcal O_\gamma(a)T^\dagger=\mathcal O_{\gamma-1}(a).
\end{eqnarray}
Hence, we introduce the following operators
\begin{eqnarray}\label{r7}
A:=\frac{1}{2}C_\gamma T^\dagger,\,\,\,\,A^\dagger:=\frac{1}{2}TC^\dagger_\gamma,\,\,\,\,N:=\frac{1}{4}C^\dagger_\gamma
C_\gamma=\frac{1}{4}h_+,
\end{eqnarray}
which lead us to the simple harmonic oscillator algebra
\begin{eqnarray}\label{r8}
[A,A^\dagger]=1,\,\,\,[N,A]=-A,\,\,\,[N,A^\dagger]=A^\dagger.
\end{eqnarray}
Therefore, the action of these operators on normalized  eigenfunctions will
be
\begin{eqnarray}\label{r9}
A\psi_{n'}^\gamma=\sqrt{n'}\psi_{n'-1}^\gamma,\,\,\,A^\dagger\psi_{n'}^\gamma=\sqrt{n'+1}\psi_{n'+1}^\gamma,\,\,\,N\psi_{n'}^\gamma=n'\psi_{n'}^\gamma.
\end{eqnarray}
The last equation in (\ref{r9}) gives $\bar E_{n'}^\gamma=4(n'-\frac{\gamma}{2}+\frac{1}{4})$
which is in agreement with the energy spectrum obtained by direct solving
of WDW in Eq. (\ref{1,5}). Let us now obtain $\alpha$ for the case of $\gamma=0$ (equivalently $q=0$ or $q=2$), for which the above generalized ladder operators reduce to the simple harmonic ladder operators. In this case, the equation $A\psi_{n'}^\gamma=\sqrt{n'}\psi_{n'-1}^\gamma$  at the
vicinity of singularity will be
 \begin{eqnarray}\label{r10}
 \left(\frac{d}{da}\psi_{n'}(a)+a\psi_{n'}(a)\right)|_{a\rightarrow0^{+}}=2\sqrt{n'}\psi(a)|_{a\rightarrow0^+}.
 \end{eqnarray}
 Now, inserting  condition  (\ref{l}) into this equation gives
 \begin{eqnarray}\label{r11}
 (\alpha+a)\psi_{n'}(a)|_{a\rightarrow0^+}=2\sqrt{n'}\psi_{n'-1}(a)|_{a\rightarrow0^+}.
 \end{eqnarray}
 The wave function in this case is an Hermite polynomial of order $n^{\prime}$
and for even values of the quantum number $n'$, $\psi_{n'}(0^+)\neq0$
 and $\psi_{n'-1}(0^+)=0$. Consequently, for the even values of $n'$, Eq. (\ref{r11})  gives $\alpha=0$. On the other hand, for the odd values of quantum number $n'$,
 $\psi_{n'}(0^+)=0$ and $\psi_{n'-1}(0^+)\neq0$, which gives $1/\alpha=0$
 or $\alpha=\infty$.
   From the above possibilities, the behavior of a given wave packet can be investigated, namely, near the singularity. Thus it can be proposed that the wave
function
vanishes at the singularity, i.e., (De Witt or Dirichlet boundary proposal)
\begin{eqnarray}\label{1,6}
\Psi(a,\widetilde{\phi})|_{a=0}=0,
\end{eqnarray}
or, as proposed in \cite{24,SHJ}, one can employ instead the Neumann boundary condition
\begin{eqnarray}\label{1,7}
\frac{\partial\Psi}{\partial a}|_{a=0}=0.
\end{eqnarray}
  Let us now return to the general case at the presence
of factor ordering, $\gamma\neq0$. Because of the simultaneous vanishing of wave function (\ref{1,5}) and
its first derivative at the singularity, $\psi^\gamma_{n'}(0^+)=0=d\psi_{n'}^\gamma/da(0^+)$,
the equation $\frac{d\psi}{da}(0^+)=\alpha\psi(0^+)$ is trivial and it does
not gives us any specific value for $\alpha$.

\subsection{Hidden symmetries, factor ordering, and Hilbert space}
The Universe is considered as a whole in quantum cosmology; i.e., there is nothing external to the
Universe. In this respect, an independent physical law may define appropriate boundary conditions \cite{26}. Or, as we
discuss herein, symmetries of the cosmological model under investigation may suggest arguments for that selection
\cite{10,11}. Indeed, from the hidden symmetries present in our model, we can extract different types of boundary conditions. To
this aim, we employ the Dirac observables of the cosmological model.
Let us be more clear. Notice that the Poisson bracket algebra associated with the sets (\ref{4,4}) and (\ref{4,5}) can be
promoted
into a $su(1,1)$ algebra \cite {24 sfd}, making use of the factor-ordering possibility that characterizes the gravitational sector of the Hamiltonian, whereas regarding the scalar field part, it remains unchanged. For the gravitational sector we can, therefore, write
\begin{eqnarray}\label{5,0}
\begin{cases}
 K_0= \frac{1}{4} \left[- \frac{d^2}{da^2} + \frac{q(q-2)}{4a^2}+a^2\right], \\
  K_{\pm}= \frac{1}{4} \left[\frac{d^2}{da^2}-\frac{q(q-2)}{4a^2}+ a^2
    \mp 2\left(a \frac{d}{da} + \frac{1}{2}\right)\right],
\end{cases}
\end{eqnarray}
with the following commutation relations:
\begin{eqnarray}\label{5,5}
\left[K_{+},K_{-}\right]=-2K_{0},\hspace{.3cm}
\left[K_{0},K_{\pm}\right]=\pm K_{\pm}.
\end{eqnarray}
The action of the above generators on a set of basis eigenvectors $|k,l\rangle$ is given by
\begin{eqnarray}\label{5,6}
\begin{cases}
K_{0}|k,l\rangle=(k+l)|k,l\rangle,\\
K_{+}|k,l\rangle=\sqrt{(2k+l)(l+1)}|k,l+1\rangle,\\
K_{-}|k,l\rangle=\sqrt{l(2k+l-1)}|k,l-1\rangle.\\
\end{cases}
\end{eqnarray}
For the scalar field part, we have \cite{10}
\begin{eqnarray}\label{5,5,2}
\left[J_{+},J_{-}\right]=-2J_{0},\hspace{.3cm}
\left[J_{0},J_{\pm}\right]=\pm J_{\pm}.
\end{eqnarray}
Defining eigenvectors $|j,m\rangle$ as the eigenvectors of $J_{0}$, the actions of the above generators are
\begin{eqnarray}\label{5,6,2}
\begin{cases}
J_{0}|j,m\rangle=(j+m)|j,m\rangle,\\
J_{+}|j,m\rangle=\sqrt{(2j+m)(m+1)}|j,m+1\rangle,\\
J_{-}|j,m\rangle=\sqrt{m(2j+m-1)}|j,m-1\rangle.
\end{cases}
\end{eqnarray}
The above commutation relations represent the Lie algebra of $su(1,1)$. The spectrum of eigenvalues of this Lie algebra
constitutes a discrete series of positive quantities and is labeled by the Bargmann indices $k$ and $j$, which are positive real numbers,
i.e., $k>0$ and $j>0$, where $m$ and $l$ are non-negative integers. Moreover, the Casimir operator for the gravitational part
is defined as \cite{22}
\begin{eqnarray}\label{5,8}
\begin{cases}
K^{2}:=K_{0}^{2}-\frac{1}{2}(K_{+}K_{-}+K_{-}K_{+}),\\
K^{2}|k,l\rangle=k(k-1)|k,l\rangle,
\end{cases}
\end{eqnarray}
with the following commutation relations
\begin{eqnarray}\label{5,9}
\left[K^{2},K_{0}\right]=0,\hspace{.3cm}
\left[K^{2},K_{\pm}\right]=0.
\end{eqnarray}
And for the scalar field part, the corresponding Casimir operator is
\begin{eqnarray}\label{5,8,2}
\begin{cases}
J^{2}:=J_{0}^{2}-\frac{1}{2}(J_{+}J_{-}+J_{-}J_{+}),\\
J^{2}|j,m\rangle=j(j-1)|j,m\rangle,
\end{cases}
\end{eqnarray}
with the commutation relations as
\begin{eqnarray}\label{5,9,2}
\left[J^{2},J_{0}\right]=0,\hspace{.3cm}
\left[J^{2},J_{\pm}\right]=0.
\end{eqnarray}
Thus, the irreducible representation of these two $su(1,1)$ Lie algebras is determined by the numbers $j$ and $k$ and the eigenstates of $\{J^{2},
K^{2}, J_{0}, K_{0}\}$. Furthermore, the Hamiltonian can be written as
\begin{eqnarray}\label{5,10}
{\ H}=4\left(K_{0}-J_{0}\right),
\end{eqnarray}
which means that the Casimir operator commutes with the Hamiltonian
\begin{eqnarray}\label{5,11}
[K^{2}, H]=0,\hspace{.3cm}
[J^{2}, H]=0.
\end{eqnarray}
Since, $K^{2}$, $K_{0}$, $J^{2}$ and $J_{0}$ commute with the Hamiltonian, they leave the physical Hilbert space $V_{H}$
invariant. Consequently, we choose $\{K_{0},K^{2},1\}$ for the gravitational section and  $\{J_{0},J^{2},1\}$ for the
scalar
field part, as physical operators of the model.

For the scalar field part, according to the definition (\ref{4,5}), the Casimir operator of $su(1,1)$ reduces to $J^{2}=j(j-1)=-\frac{3}{16}$. Hence, the Bargmann index $j=\{\frac{1}{4},\frac{3}{4}\}$ is
a gauge-invariant observable of the quantum cosmological model. Furthermore, from (\ref{1,2}), (\ref{5,6,2}) and
(\ref{5,10}) we obtain
\begin{eqnarray}\label{5,12}
E_{n}=4(j+m).
\end{eqnarray}
Therefore, the scalar field sector of the Hilbert space, by means of the Hamiltonian constraint, can be classified in terms of the Bargmann index, allowing us to
establish two invariant odd and even subspaces:
\begin{eqnarray}\label{5,13}
\begin{cases}
E_{j=\frac{3}{4},m}=2\left(2m+1+\frac{1}{2}\right),~~~~~~
V_{H_{\widetilde{\phi}},j=\frac{3}{4}}=\{|\frac{3}{4},m\rangle\},\\
E_{j=\frac{1}{4},m}=2\left(2m+\frac{1}{2}\right),~~~~~~~~~~~ V_{H_{\widetilde{\phi}},j=\frac{1}{4}}=\{|\frac{1}{4},m\rangle\}.
\end{cases}
\end{eqnarray}
Similarly for the gravitational sector: using definitions (\ref{5,8}), the Casimir operator of the gravitational part reduces identically to $K^{2}=k(k-1)=\frac{1}{16}\left(q+1\right)\left(q-3\right)$. Thus the Bargmann index
$k=\{\frac{1}{2}-\frac{1}{4}|1-q|,\frac{1}{2}+\frac{1}{4}|1-q|\}$ is a  gauge-invariant observable of the quantum
cosmological model. The Bargmann index must be positive and real valued, which restricts $q$ to lie in the interval
\begin{eqnarray}\label{5,20}
-1\leqslant q\leqslant3.
\end{eqnarray}
Thus, the factor-ordering parameter $q$ is determined through the Bargmann indices, which are observables of our model.
In particular, $q=1$ is the covariant ordering used by Isham \cite{Isham:1992ms}. Although there exists an infinity of possibilities of factor ordering regarding (\ref{1,0}), the symmetries of the model constrain the values for factor ordering as explored above.

In addition, from (\ref{1,5}), (\ref{5,6,2}) and
(\ref{5,10}) we obtain
\begin{eqnarray}\label{5,12,a}
\bar{E}_{n^{\prime}}^\gamma=4(k+l).
\end{eqnarray}

Moreover,  the gravitational sector of the Hilbert space can be classified in terms of the Bargmann index as
\begin{eqnarray}\label{5,14}
\begin{cases}
\bar E_{n',q}=4(l+\frac{1}{2}-\frac{1}{4}|1-q|),\,\,\,\,V_{H_a,k=\frac{1}{2}-\frac{1}{4}|1-q|}=|\frac{1}{2}-\frac{1}{4}|1-q|,l\rangle,\\
E_{n',q}=4(l+\frac{1}{2}+\frac{1}{4}|1-q|),\,\,\,\,V_{H_a,k=\frac{1}{2}+\frac{1}{4}|1-q|}=|\frac{1}{2}+\frac{1}{4}|1-q|,l\rangle.
\end{cases}
\end{eqnarray}

The states of the Hilbert space can be classified as
\begin{eqnarray}\label{5,17}
V_{H=0}=V_{H_{a}}\otimes\ V_{H_{\widetilde{\phi}}}.
\end{eqnarray}
Therefore, the gauge invariance of the Bargmann indices implies a partition of the Hilbert space into four disjointed
invariant subspaces.

\section{SUMMARY AND DISCUSSION}
In general, the wave function retrieved from the WDW equation with appropriate boundary conditions should describe the Universe. An
interesting approach has been provided in \cite{10,11}, where boundary proposals can be selected by means of a careful
analysis of the algebra associated with the Dirac observables. In Ref. \cite{10}, a closed FLRW universe filled with either
dust or radiation
was considered, in which, the Hamiltonian of that model is equivalent to a one-dimensional simple harmonic oscillator. The
$su(1,1)$ hidden symmetry of that model with the set of gauge-invariant of the Bargmann values
$\{\frac{1}{4},\frac{3}{4}\}$ split the underlying Hilbert space into two disjoint invariant subspaces. These subspaces
were shown to be corresponding to different choice of boundary conditions. In Ref. \cite{11},
with
a similar procedure, the hidden symmetries present in a pre-big bang model were identified, namely, $u(1,1)$  together with time reversal and
parity. These lead to the Hamiltonian of the model being equivalent to an oscillator-ghost-oscillator system.
The two-mode realization of the $su(1,1)$ algebra, together with the Hamiltonian constraint implied a degenerate Bargmann
index. However, the scale factor duality of that model plus time reversal, still allowed boundary conditions to be selected.

In this paper, we considered a conformally coupled scalar field in a closed Friedmann universe. The WDW
equation is separated into a scalar field part plus the gravitational sector. We made use of the corresponding phase
space quantization of the Casimir operator, as an operator which commutes with the Hamiltonian. We further showed that the Bargmann
indices are gauge-invariant observables of the quantum cosmological model.
From the vanishing of the commutator of the $su(1,1)$ generator with the Hamiltonian of the system, in addition to the
gauge
invariance of the Bargmann indices, we found it possible to select the wave function of the Universe. In other words, our
proposed framework \cite{10,11} applied to the model in Sec II, implied a specific set of boundary conditions, to which a
selection of the factor ordering as an observable was also admissible.

More concretely, the Hamiltonian of our model consists of a one-dimensional simple harmonic oscillator for the scalar field part and a one-dimensional simple harmonic oscillator an inverse square potential arose from the factor ordering for the gravity sector. The
Hamiltonian has the $su(1,1)$ hidden symmetry with the set of gauge-invariant Bargmann values $\{\frac{1}{4},\frac{3}{4}\}$
and $\{\frac{1}{2}(1+\frac{1}{2}|1-q|),\frac{1}{2}(1-\frac{1}{2}|1-q|)\}$ for the scalar field part and the gravitational sector, respectively. These split the underlying Hilbert space into four disjoint invariant subspaces. The factor-ordering parameter $q$ is subsequently specified through the admissible gauge invariant Bargmann indices.

Finally, we must emphasize the following:
\begin{enumerate}
\item
Our results are retrieved on a very restrictive scenario: a homogeneous and isotropic cosmological model, with a very particular coupling between gravity and matter, which enables those sectors to be separated in the WDW equation (cf. Sec. III). It would be interesting to establish if a similar (or at least somewhat related) intertwining pattern emerges in other models (e.g., within a minisuperspace, in particular with perturbation modes obtained from $\widetilde{\phi}$).
\item
The issues of initial condition choice and Dirac observables, have been discovered in \cite{Craig:2012zz,Ashtekar:2007em}.The problem of time is analyzed through relational observables in the model based on constructing the decoherence functional for WDW quantization \cite{Craig:2012zz}. We hope to extend the analysis presented here and explore the implication of our analysis within the results of \cite{Craig:2012zz,Ashtekar:2007em} in future works. Specifically, it may be worthy to investigate the similarities between Dirac observables presented in Eqs. (\ref{5,0})-(\ref{5,8,2}) with respect to the (relational) Dirac observables presented in \cite{Ashtekar:2007em}.
\item
If the hidden symmetries of the full WDW equation for quantum gravity could be established, this would allow us to extend the framework in this paper towards a wider context. Although the underlying approach in our paper is tied to concrete models bearing characteristic symmetries, it may nevertheless suggest valuable insights toward discussing the WDW equation in broader settings.
\item
Finally, it is curious that within our analysis the issue of shape invariance, which is a feature present in some approaches of supersymmetric quantum mechanics \cite{Cooper:1994eh}, has emerged to be employed. Although bearing in mind the gap between supersymmetric field theories \cite{Dine:2007zp} and supersymmetric quantum mechanics \cite{Cooper:1994eh} (a somewhat reduced toy model for the former), if any evidence is advanced in the future to support supersymmetry, then it will be interesting to consider how boundary conditions, factor-ordering, and hidden symmetries become intertwined in a manner that a supersymmetry somehow can be made to appear in the discussion.
\end{enumerate}

\textbf{Acknowledgments} This work was supported in part by the
grant PEst-OE/MAT/UI0212/2014.


\end{document}